\documentclass
[superscriptaddress,secnumarabic,amssymb,amsmath,nobibnotes,aps,prd,showkeys,nofootinbib,nopacsnumber,onecolumn,10pt]{revtex4}%
\usepackage{graphicx}
\usepackage{bm}
\usepackage{amsmath}
\usepackage{amsfonts}
\usepackage{amssymb}
\usepackage[utf8]{inputenc}%
\providecommand{\U}[1]{\protect\rule{.1in}{.1in}}

\newcommand{\be}{\begin{equation}}
\newcommand{\ee}{\end{equation}}

\newcommand{\mincir}{\raise
-3.truept\hbox{\rlap{\hbox{$\sim$}}\raise4.truept\hbox{$<$}\ }}
\newcommand{\magcir}{\raise
-3.truept\hbox{\rlap{\hbox{$\sim$}}\raise4.truept\hbox{$>$}\ }}

\begin{document}

\title{Asymptotic Solutions of Radiating Stars}

\author{R. S. Bogadi}
\email{robertb@dut.ac.za}
\affiliation{Department of Mathematics, Faculty of Applied Sciences, Durban University of
Technology, Durban 4000, South Africa}
\author{G. Leon}
\email{genly.leon@ucn.cl}
\affiliation{Departamento de Matem\`{a}ticas, Universidad Cat\`{o}lica del Norte, Avda.
Angamos 0610, Casilla 1280 Antofagasta, Chile}
\affiliation{Institute of Systems Science, Faculty of Applied Sciences, Durban University of
Technology, Durban 4000, South Africa}
\author{M. Govender}
\email{megandhreng@dut.ac.za}
\affiliation{Department of Mathematics, Faculty of Applied Sciences, Durban University of
Technology, Durban 4000, South Africa}
\author{K.S. Govinder}
\email{govinder@ukzn.ac.za}
\affiliation{Astrophysics Research Centre, Discipline of Mathematics, School of Agriculture and Science, University of KwaZulu-Natal, Private Bag X54001, 4000 Durban,
South Africa}
\author{S. Maharaj}
\email{maharaj@ukzn.ac.za}
\affiliation{Astrophysics Research Centre, Discipline of Mathematics, School of Agriculture and Science, University of KwaZulu-Natal, Private Bag X54001, 4000 Durban,
South Africa}
\author{A. Paliathanasis}
\email{anpaliat@phys.uoa.gr}
\affiliation{Institute of Systems Science, Faculty of Applied Sciences, Durban University of
Technology, Durban 4000, South Africa}
\affiliation{Departamento de Matem\`{a}ticas, Universidad Cat\`{o}lica del Norte, Avda.
Angamos 0610, Casilla 1280 Antofagasta, Chile}
\affiliation{National Institute for Theoretical and Computational Sciences (NITheCS), South Africa.}

\begin{abstract}
We investigate the evolution of the surface of radiating stars by studying the asymptotic behaviour of exact solutions initiated via the stationary boundary condition. This boundary condition leads to a master equation in the form of a second-order nonlinear differential equation that describes the evolution of the scale factor. We examine this master equation by introducing a set of dimensionless dynamical variables, motivated by similar approaches in cosmological settings. We derive the stationary points of the system in the presence of charge and a cosmological constant. Furthermore, we construct criteria for the initial conditions in order that the asymptotic limit approaches a static geometry.

\end{abstract}
\date{\today}
\maketitle

\section{Introduction}

The problem of radiative, gravitational collapse of stellar matter is an ideal scenario and most evident natural phenomenon for the application of Einstein's theory of General Relativity (GR). Since the pioneering work of Oppenheimer and Snyder \cite{opp}, models have greatly improved by using more realistic interior metrics incorporating variable matter distributions and more generalised matter fields allowing for the incorporation of anisotropies and other non-perfect fluid attributes. Notable contributions to the modelling of gravitational collapse are those by de Oliveira et al.\cite{oli}, Bonnor et al.\cite{bon}, Herrera et al.\cite{her1}, Chan et al.\cite{chan}, Ivanov \cite{ivan} and Pretel et al. \cite{pre1,pre2}. More sophisticated modelling has required consideration of the origin of local anisotropy and cracking due to tidal forces \cite{her3}, heat transport processes and causality \cite{her2} and more recently, the concept of complexity \cite{her4,bog2}. There are thus now numerous solutions to the field equations that describe a relativistic radiating star, having been obtained by closing the system of coupled field equations in certain, physically motivated ways. Restrictions may be placed on the metric potentials, the kinematical quantities and/or the matter variables such as pressure isotropy and equations of state. Initial static configurations may be obtained using realistic equations of state \cite{pre2} and imposing an equation of state during the collapse process has also been attempted \cite{bog3}.

Of primary importance in modelling gravitational collapse is the application of the appropriate boundary condition as put forward by Santos \cite{san}. This results in a second-order, non-linear differential equation which is not easily solved analytically. The situation is further compounded by incorporating the additional attributes such as non-perfect fluid characteristics and electric charge. The boundary condition has been successfully solved for shear-free systems in which the metric potentials are variable separable in space and time \cite{bon}, and also for systems in higher dimensions \cite{bui,bog}. For metric functions where this is not the case, more systematic mathematical tools are required, such as the quantitative Lie group theoretic method \cite{pal2,nai}. The Lie approach has been applied to Euclidean, geodesic and conformally flat stellar models as well as shear-free relativistic models and geometry embedding models such as the Karmarkar condition \cite{pal2,mah2}. In higher dimensions, the boundary condition is even more cumbersome and Lie symmetries assist in reducing the order of the differential equation \cite{nai}.  

The basic form of the heat-flux boundary condition has also appeared in modified gravity theories and higher dimensional studies. Even in standard four-dimensional GR, an explicit, analytical solution is not easily obtained and the Lie group theoretic approach is typically used in finding solutions. This was done for the general spherically symmetric spacetime metric \cite{nai}. A solution for a shear-free metric in higher dimensions has also been obtained implicitly as an infinite series solution \cite{bog}. 

Another less common, but equally important method which may be employed in analysing the boundary condition is qualitative phase plane analysis. This allows us to determine the nature of the asymptotic end state of the collapsing medium. A notable study is that of Leon et al. \cite{leo} which considered the global dynamics and asymptotic behaviour of a particular radiating model. 

In this study, we consider a metric involving both charge and the cosmological constant. The cosmological constant has been considered in other studies \cite{mah,thi,you} as it could play a role in the stability of a dynamical process. The inclusion of charge, encapsulated in the Reissner–Nordstr\"{o}m metric, is especially important as electrostatic forces can be comparable to gravitational forces in the early stages of collapse, and perhaps prevent a self-gravitating collapse from proceeding \cite{bon2}. The Reissner–Nordstr\"{o}m static solution \cite{rei} was discovered almost concurrently with the Schwarzschild solutions and early development of General Relativity. 

{Nevertheless, in \cite{mah}, the asymptotic behaviours that were derived correspond to static solutions. In contrast, in the present study we adopt a different approach. Specifically, we rewrite the boundary condition in the form of a Friedmann-like equation and introduce a normalization of the parameters in terms of dimensionless variables. The resulting reduced dynamical system is then compactified using Poincar\'{e} variables, and the phase-plane trajectories are constructed and analysed. Within this framework, we are able to derive the analytic form of the evolving spacetime in the asymptotic regime. This method provides deeper insight into the physical behaviour of the spacetime geometry, as also reveal important information regarding the initial value problem.}


\section{Radiating Stars}

\label{sec2}

We make use of a shear-free, spherically symmetric geometry which is suitable for modelling the collapse of dust or fluid spheres. The line element is then given by

\begin{equation}
ds^{2}=-A^{2}\left(  r\right)  dt^{2}+B^{2}\left( t\right) \left(dr^{2}+r^{2}d\Omega^{2}\right),
\label{lm1}
\end{equation}
where $A$ and $B$ are the metric functions or gravitational potentials as sometimes referred to, and are to be determined via physical and/or geometrical constraints. This shear-free metric was studied in detail by Bonnor et al. \cite{bon}. 
We consider matter that may also be electrically charged. {The line element (\ref{lm1}) describes the evolution of the radiating star in a nonstatic cosmological background.}

The energy-momentum tensor for charged matter which dissipates heat to the exterior is given by 

\begin{equation} \label{2} 
T_{ab} = (\rho + p_r) \, u_au_b + p_t g_{ab} + (p_r - p_t)\chi_a\chi_b + q_au_b + q_bu_a + E_{ab},
\end{equation} 
where $\rho$ is the energy density, $p_r$ and $p_t$ are radial and tangential pressures respectively. The electromagnetic stress-energy tensor is given by 

\begin{equation}
E_{ab} = \frac{1}{4\pi}\left(F_a^cF_{bc} - \frac{1}{4}g_{ab}F_{cd}F^{cd}\right),
\end{equation}
where $F_{ab}$ is the electromagnetic field tensor.

The heat flow vector $q^a$ is orthogonal to the velocity vector so that $q^a u_a = 0$. 
The Einstein field equations are given by

\begin{equation}
R_{ab} - \frac{1}{2}R g_{ab} + \Lambda g_{ab} = 8\pi T_{ab}~,
\end{equation}
where $R_{ab}$ and $R$ are the Ricci curvature tensor and scalar curvature respectively, and $\Lambda$ is the cosmological constant. More explicitly, the matter variables are given by

\begin{eqnarray}   \label{t4}
8\pi \rho &=& \frac{3}{A^2}\frac{{\dot{B}}^2}{B^2} - \frac{1}{B^2}\left(\frac{2B''}{B} - \frac{{B'}^2}{B^2} + \frac{4}{r}\frac{B'}{B} \right) - \frac{l^2}{r^4 B^4} - \Lambda~,  \label{t4a} \\ 
8\pi p_r &=& \frac{1}{A^2} \left(-\frac{2\ddot{B}}{B} - \frac{{\dot{B}}^2}{B^2} + 2\frac{\dot{A}}{A}\frac{\dot{B}}{B} \right)  \nonumber \\
&& + \frac{1}{B^2} \left(\frac{{B'}^2}{B^2} + 2\frac{A'}{A}\frac{B'}{B} + \frac{2}{r}\frac{A'}{A} + \frac{2}{r}\frac{B'}{B} \right) + \frac{l^2}{r^4 B^4} + \Lambda~,  \label{t4b}  \\ 
8\pi p_t &=& \frac{1}{A^2}\left(-\frac{2\ddot{B}}{B} - \frac{{\dot{B}}^2}{B^2} + 2\frac{\dot{A}}{A}\frac{\dot{B}}{B} \right) \nonumber \\
&& + \frac{1}{B^2}\left(- \frac{{B'}^2}{B^2} + \frac{1}{r}\frac{A'}{A} +  \frac{1}{r}\frac{B'}{B} + \frac{A''}{A} + \frac{B''}{B}\right) - \frac{l^2}{r^4 B^4} + \Lambda~, \label{t4c}  \\ \nonumber \\
8\pi q &=& \frac{2}{B^2} \left(\frac{\dot{B}}{AB}\right)' ~, \label{t4d} \\
4\pi\eta &=& \frac{l'}{r^2B^3}~,
\end{eqnarray}
where now the dot and prime are derivatives with respect to time and radial distance respectively. The presence of $E_{ab}$ in the energy-momentum tensor gives an equation for the proper charge density $\eta$ with $l(r)$ being the charge contained within a sphere of radius $r$. \\

Matching of the interior spacetime with a radiation filled exterior metric across a hypersurface $\Sigma$ which is the surface of the star is needed for completing the model. An exterior, Vaidya-Bonner-de Sitter metric is used \cite{sha}, given by

\begin{eqnarray}
ds^2 &=& -\left(1 - \frac{2m(\upsilon)}{\text{r}} + \frac{Q^2(\upsilon)}{\text{r}^2} - \frac{\Lambda \text{r}^2}{3}\right)d\upsilon^2  \nonumber \\
&& - 2d\upsilon d\text{r} + \text{r}^2\left(d\theta^2 + \sin^2\theta~ d\phi^2 \right),
\end{eqnarray}
in which $Q$ is the charge $Q = l(r)|_\Sigma$. Matching of the interior and exterior metrics then results in the boundary condition,

\begin{equation}
p_r|_\Sigma = (q B)|_\Sigma~.
\end{equation}

Upon substitution of the matter variables (\ref{t4b}) and (\ref{t4d}, we obtain the second-order, non-linear differential equation,

\begin{equation} \label{BC}
2B\ddot{B}+\dot{B}^{2}+\alpha\dot{B}+\beta+\frac{\gamma}{B^{2}}+\delta B^{2} = 0~,
\end{equation}

where \{$\alpha = -2A' ~,~ \beta = -2A'A/r ~,~ \gamma = -l^2A^2/r^4  ~,~  \delta = -\Lambda A^2$\}, all evaluated at the surface boundary $\Sigma$.

This equation then dictates the temporal behaviour of the collapse process. Similar equations have appeared in many recent articles on non-adiabatic, self-gravitating collapsing systems \cite{pal1,mah}.

We now define a new quantity $H$ such that

\begin{equation}
\dot{B} = BH~
\end{equation}
which then helps to transform (\ref{BC}) to first order, namely,

\begin{equation}
\left(2\frac{\dot{H}}{H^{2}}+3\right) = -\frac{\alpha}{BH}-\frac{1}{H^{2}}\left( \frac{\beta}{B^{2}}+\frac{\gamma}{B^{4}}+\delta\right)~.
\label{pp1}
\end{equation}

{This takes the form of the second Friedman equation with the pressure components of fluids described by curvature $\frac{\beta}{B^{2}},~$radiation $\frac{\gamma}{B^{4}}$, cosmological constant, and a new component $\frac{\alpha}{BH}$. We shall refer to this equation as the master equation.} 

{The differential equation \ref{pp1}, describes the evolution of the cosmological background space, where the radiation stars lies. Thus, it's solution corresponds to the background geometry.}

\section{Asymptotic analysis}

\label{sec3}

We continue our analysis by investigating the asymptotic dynamical behaviour of the master equation by making use of dimensionless variables in the $H$-normalization approach,

\[
\Omega_{a}=\frac{\alpha}{BH},~\Omega_{\beta}=\frac{\beta}{B^{2}H^{2}}%
~,~\Omega_{\gamma}=\frac{\gamma}{B^{4}H^{2}},~\Omega_{\delta}=\frac{\delta}{H^{2}}.
\]

{We use the notation used in cosmology, where now $\Omega_{\alpha}$,~$\Omega_{\beta}$,~$\Omega_{\gamma}$ and $\Omega_{\delta}$ describe the \textquotedblleft energy densities\textquotedblright\ for the fluid components with coefficient $\alpha$, the curvature-like term with coefficient $\beta$, the charge with coefficient $\gamma$ and the cosmological constant with coefficient $\delta$. We assume that $H\neq0$. Moreover, by definition, the only constraint on these variables is for $\Omega_{\gamma}\geq0$ since $\gamma\geq0$.}

A stationary point for the master equations follows when the dimensionless variables $\Omega_{\alpha}$,~$\Omega_{\beta}$,~$\Omega_{\gamma}$ and $\Omega_{\delta}$ are constant. Indeed, we derive the evolution equations for these dynamical variables, and then write the master equation in the equivalent system. We obtain the dynamical system,

\begin{align}
\frac{d\Omega_{\alpha}}{d\tau} &  =\frac{1}{2}\Omega_{\alpha}\left(1+\Omega_{\alpha}+\Omega_{\beta}+\Omega_{\gamma}+\Omega_{\delta}\right),\label{c.01}\\
\frac{d\Omega_{\beta}}{d\tau} &  =\Omega_{\beta}\left(  1+\Omega_{\alpha}+\Omega_{\beta}+\Omega_{\gamma}+\Omega_{\delta}\right)  ,\label{c.02}\\
\frac{d\Omega_{\gamma}}{d\tau} &  =\Omega_{\gamma}\left(  -1+\Omega_{\alpha}+\Omega_{\beta}+\Omega_{\gamma}+\Omega_{\delta}\right)  ,\label{c.03}\\
\frac{d\Omega_{\delta}}{d\tau} &  =\Omega_{\delta}\left(  3+\Omega_{\alpha}+\Omega_{\beta}+\Omega_{\gamma}+\Omega_{\delta}\right)  \label{c.04}%
\end{align}

\qquad where the master equation now reads%

\begin{equation}
\dot{H}=-\frac{1}{2}H^{2}\left(  3+\Omega_{\alpha}+\Omega_{\beta}%
+\Omega_{\gamma}+\Omega_{\delta}\right)
\end{equation}

and the new independent variable $\tau$ is defined as $\tau=\ln B$. 

For the asymptotic analysis we consider three different gravitational scenarios; Case A: there is no charge or cosmological constant; Case B: there exists only charge; and Case C: the system evolves in the presence of charge and the cosmological constant.

\subsection{Case A: $~A^{\prime}\neq0,~\gamma=0,~\delta=0$}

In the absence of charge and the cosmological constant, the dynamical system is reduced to the following two-dimensional system%

\begin{align}
\frac{d\Omega_{\alpha}}{d\tau}  &  =\frac{1}{2}\Omega_{\alpha}\left(
1+\Omega_{\alpha}+\Omega_{\beta}\right)  ,\label{ca.01}\\
\frac{d\Omega_{\beta}}{d\tau}  &  =\Omega_{\beta}\left(  1+\Omega_{\alpha
}+\Omega_{\beta}\right)  , \label{ca.02}%
\end{align}

where now the Hubble function is given by the expression

\begin{equation}
\dot{H}=-\frac{1}{2}H^{2}\left(  3+\Omega_{\alpha}+\Omega_{\beta}\right)~.
\label{ca.03}%
\end{equation}

The stationary points $P_{A}=\left( \Omega_{\alpha}\left(  P_{A}\right),\Omega_{\beta}\left(  P_{A}\right) \right) ~$ of the dynamical system (\ref{ca.01}), (\ref{ca.02}) are

\begin{align*}
P_{A}^{1}  &  =\left(  0,0\right)  ,~\\
P_{A}^{2}  &  =\left(  \Omega_{\alpha}\left(  P_{A}^{2}\right)  ,-1-\Omega
_{\alpha}\left(  P_{A}^{2}\right)  \right)~,
\end{align*}

where for point $P_{A}^{2}$, parameter $\Omega_{\alpha}^{0}$ is arbitrary.

The corresponding asymptotic solutions for expansion rates are calculated
$H\left(  P_{A}^{1}\right)  =\frac{3}{2\left(  t-t_{0}\right)  }$ and $H\left(  P_{A}^{2}\right)  =\frac{1}{\left(  t-t_{0}\right)  }$, thus the exact solutions for the scale factor $B\left(  t\right)  $ are $B\left(P_{A}^{1}\right)  =B_{0}\left(  t-t_{0}\right)  ^{\frac{2}{3}}$ and $B\left(P_{A}^{2}\right)  =B_{0}\left(  t-t_{0}\right)$. {Recall that these are the asymptotic solutions for the cosmological background.}

As far as the stability properties are concerned, we find that the resulting eigenvalues for point $P_{A}^{1}$ are $\left\{  1,\frac{1}{2}\right\}  $, which means that the stationary point is a source, and for $P_{A}^{2}$ we calculate $\left\{  0,-1-\frac{\Omega_{\alpha}^{0}}{2}\right\} $. Due to the zero eigenvalue, the Center Manifold Theorem (CMT) should be applied in order to infer about the nature of the stability properties of the asymptotic solution. \ We remark that the line $1+\Omega_{\alpha}+\Omega_{\beta}=0$, \ describes the family of points $P_{A}^{2}$, and since it multiplies both equations the local center manifold is every point on the algebraic constraint. Hence, for $\Omega_{\alpha}^{0}>-2$ the points are attractors.

\subsubsection{Poincar\'{e} Variables}

The dynamical system (\ref{ca.01}), (\ref{ca.02}) is however not compactified and the solution trajectories can reach infinity. In order to investigate the existence of stationary points at infinity, we introduce the Poincar\'{e} map%

\begin{equation}
\Omega_{a}=\frac{X}{\sqrt{1-X^{2}-Y^{2}}},~\Omega_{\beta}=\frac{Y}%
{\sqrt{1-X^{2}-Y^{2}}}\text{,~}d\tau=\sqrt{1-X^{2}-Y^{2}}dT\text{.}%
\end{equation}

Hence, system (\ref{ca.01}), (\ref{ca.02}) reads%

\begin{align}
\frac{dX}{dT}  &  =-\frac{1}{2}X\left(  X^{2}+2Y^{2}-1\right)  \left(
X+Y+\sqrt{1-X^{2}-Y^{2}}\right)  ,\\
\frac{dY}{dT}  &  =-\frac{1}{2}Y\left(  X^{2}+2Y^{2}-1\right)  \left(
X+Y+\sqrt{1-X^{2}-Y^{2}}\right)  .
\end{align}

The stationary points $Q=\left(  X\left(  Q\right)  ,Y\left(  Q\right)\right) ~$at infinity satisfy the additional constraint $1-X^{2}-Y^{2}=0$; thus the stationary points are calculated as%

\begin{align*}
Q_{A}^{1\pm}  &  =\left(  \pm1,0\right)  ,~\\
Q_{A}^{2\pm}  &  =\left(  0,\pm1\right)  ,\\
Q_{A}^{3\pm}  &  =\pm\frac{\sqrt{2}}{2}\left(  1,-1\right)  .
\end{align*}

Points $Q_{A}^{1\pm}$ have the eigenvalues $\left\{ -1,\frac{1}{2}\right\} $ and $\left\{  1,-\frac{1}{2}\right\} $ respectively, from which we infer that they are saddle points. On the other hand, for points $Q_{A}^{2\pm}$ we calculate $\left\{ -2,-\frac{1}{2}\right\}  $ and $\left\{  2,\frac{1}{2}\right\} $, that is, $Q_{A}^{2+}$ is an attractor and $Q_{A}^{2-}$ is a source. Finally, points $Q_{A}^{3\pm}$ are always saddle points.

As far as the physical properties of the attractor $Q_{A}^{2+}$ are concerned, we calculate $H\left( t\right) \rightarrow0$, which corresponds to $B\left( t\right) \rightarrow B_{0}$, so that the static solution is recovered at infinity. Similar physical properties are encountered for the point $Q_{A}^{1+}$. On the other hand, we find that $Q_{A}^{1-}$ and $Q_{A}^{2-}$ describe Big Rip scenarios, $H\left(  t\right) \rightarrow\infty$. Finally, for the points $Q_{A}^{3\pm}$ we find that the asymptotic behaviour for the scale factor $B_{0}\left(t - t_{0}\right) ^{\frac{2}{3}}$ is similar to that of $P_{A}^{1}$.

\begin{figure}[ptbh]
\centering\includegraphics[width=0.5\textwidth]{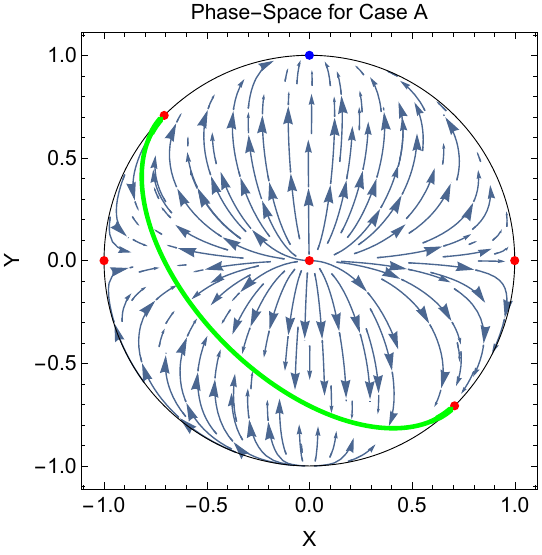}\caption{Phase-Space for the dynamical system (\ref{ca.01}), (\ref{ca.02}) in the compactified variables. The unstable asymptotic solutions are marked with red points, the attractor at the infinity is marked with blue, and the green line describes the family of points $P_{A}^{2}$. }%
\label{fig1}%
\end{figure}

The above results are summarized in Table \ref{table1}. In Fig. \ref{fig1} we present the phase-space for the dynamical system (\ref{ca.01}), (\ref{ca.02}) within the compactified variables.

\begin{table}[tbp] \centering
\caption{Case A: Stationary points and stability properties}\label{table1}
\begin{tabular}
[c]{ccc}\hline\hline
\textbf{Point} & \textbf{Stability} & $\mathbf{B}\left(  t\right)  $\\\hline
$P_{A}^{1}$ & Source & $B_{0}\left(  t-t_{0}\right)  ^{\frac{2}{3}}$\\
$P_{A}^{2}$ & Attractor $\Omega_{\alpha}>-2$ & $B_{0}\left(  t-t_{0}\right)
$\\
$Q_{A}^{1-}$ & Saddle & $B_{0}$\\
$Q_{A}^{1-}$ & Saddle & Big Rip\\
$Q_{A}^{2+}$ & Attractor & $B_{0}$\\
$Q_{A}^{2-}$ & Source & Big Rip\\
$Q_{A}^{3+}$ & Saddle & $B_{0}\left(  t-t_{0}\right)  ^{\frac{2}{3}}$\\
$Q_{A}^{3-}$ & Saddle & $B_{0}\left(  t-t_{0}\right)  ^{\frac{3}{2}}%
$\\\hline\hline
\end{tabular}
\end{table}%

\subsection{Case B: $~A^{\prime}\neq0,~\gamma\neq0,~\delta=0$}

In the presence of the charge, the dynamical system reads%

\begin{align}
\frac{d\Omega_{\alpha}}{d\tau} &  =\frac{1}{2}\Omega_{\alpha}\left(
1+\Omega_{\alpha}+\Omega_{\beta}+\Omega_{\gamma}\right)  ,\label{cb.01}\\
\frac{d\Omega_{\beta}}{d\tau} &  =\Omega_{\beta}\left(  1+\Omega_{\alpha
}+\Omega_{\beta}+\Omega_{\gamma}\right)  ,\label{cb.02}\\
\frac{d\Omega_{\gamma}}{d\tau} &  =\Omega_{\gamma}\left(  -1+\Omega_{\alpha
}+\Omega_{\beta}+\Omega_{\gamma}\right)  ,\label{cb.03}%
\end{align}

and the master equation becomes

\begin{equation}
\dot{H}=-\frac{1}{2}H^{2}\left(  3+\Omega_{\alpha}+\Omega_{\beta} + \Omega_{\gamma}\right)  .
\end{equation}

The stationary points $P_{B}=\left(  \Omega_{\alpha}\left( P\right),\Omega_{\beta}\left(  P\right)  ,\Omega_{\gamma}\left(  P\right)  \right) $ of the dynamical system (\ref{cb.01}), (\ref{cb.02}) and (\ref{cb.03}) at the finite regime are

\begin{align*}
P_{B}^{1} &  =\left(  0,0,0\right)  ,\\
P_{B}^{2} &  =\left(  \Omega_{\alpha}^{0},-1-\Omega_{\alpha}^{0},0\right)  \\
P_{B}^{3} &  =\left(  0,0,1\right)  .
\end{align*}

The stationary point $P_{B}^{3}$ has $\Omega_{\gamma}>0$, which is not physically acceptable as we require real values for the charge. Furthermore, the corresponding eigenvalues around the stationary points are calculated $P_{B}^{1}:\left(  1,\frac{1}{2},-1\right)$,~$P_{B}^{2}:\left(  0,-1-\frac{\Omega_{\alpha}^{0}}{2},-2\right)  $ and $P_{B}^{3}:\left(  2,1,1\right)  $. Thus, $P_{B}^{1}$ is a source point, $P_{B}^{3}$ is an attractor and for $P_{B}^{2}$ the application of the CMT indicates that $P_{B}^{2}$ describes a line of source points for $\Omega_{\alpha}^{0}>-2$.

\subsubsection{Poincar\'{e} Variables}

In order to study the evolution of the trajectories at the infinity regime, we apply a Poincar\'{e} map by introducing the compactified variables

\begin{align}
\Omega_{a} &  =\frac{X}{\sqrt{1-X^{2}-Y^{2}-Z^{2}}},~\Omega_{\beta}=\frac
{Y}{\sqrt{1-X^{2}-Y^{2}-Z^{2}}}\text{,~}\\
\Omega_{\gamma} &  =\frac{Z}{\sqrt{1-X^{2}-Y^{2}-Z^{2}}},~d\tau=\sqrt
{1-X^{2}-Y^{2}-Z^{2}}dT,
\end{align}

At infinity, that is, on the surface $1-X^{2}-Y^{2}-Z^{2}=0$, the stationary points of the dynamical system (\ref{cb.01}), (\ref{cb.02}) and (\ref{cb.03})  $Q=\left(  X\left(  Q\right)  ,Y\left(  Q\right)  ,Z\left(Q\right)  \right) ~$are%

\begin{align*}
Q_{B}^{1\pm} &  =\left(  \pm1,0,0\right) ~,\\
Q_{B}^{2\pm} &  =\left(  0,\pm1,0\right) ~,\\
Q_{B}^{3\pm} &  =\pm\frac{\sqrt{2}}{2}\left(  1,-1,0\right) ~,\\
Q_{B}^{4} &  =\frac{1}{2}\left(  2X_{0},-X_{0}+\sqrt{2-3X_{0}^{2}}%
,-X_{0}-\sqrt{2-3X_{0}^{2}}\right) ~,\\
Q_{B}^{5} &  =\left(  0,Y_{0},-\sqrt{1-Y_{0}^{2}}\right)~.
\end{align*}

Points, $Q_{B}^{1\pm}$,~$Q_{B}^{2\pm}$ and $Q_{B}^{3\pm}$ recover the physical properties described before in Case A.  The new stationary points $Q_{B}^{4}$ and $Q_{B}^{5}$ describe solutions with a nonzero contribution of charge in the physical space.

The asymptotic solution at the family of points $Q_{B}^{4}$ describes a power-law scale factor $B\left(  Q_{B}^{4}\right)  =B_{0}\left(t-t_{0}\right)  ^{\frac{2}{3}}$, similar to the points $Q_{B}^{3\pm}$. On the other hand, point $Q_{B}^{5}$ describes an asymptotic static solution $B\left( Q_{B}^{5}\right)  \rightarrow B_{0}$, for $\frac{\sqrt{2}}{2}<Y<1$,
otherwise, it describes a Big Rip singularity, where $B\left(  Q_{B}^{5}\right)  \rightarrow+\infty$. \ 

As far as the stability is concerned, the stationary points $Q_{B}^{1\pm},~Q_{B}^{2\pm}$ and $Q_{B}^{3\pm}$ have the same stability properties with
Case A, while points $Q_{B}^{4}$ and $Q_{B}^{5}$ describe unstable solutions,
specifically they are Saddle points. 

The above results are summarized in Table \ref{table2}. %

\begin{table}[tbp] \centering
\caption{Case B: Stationary points and stability properties}\label{table2}
\begin{tabular}
[c]{ccc}\hline\hline
\textbf{Point} & \textbf{Stability} & $\mathbf{B}\left(  t\right)  $\\\hline
$P_{B}^{1}$ & Source & $B_{0}\left(  t-t_{0}\right)  ^{\frac{2}{3}}$\\
$P_{B}^{2}$ & Attractor $\Omega_{\alpha}>-2$ & $B_{0}\left(  t-t_{0}\right)$\\
$P_{B}^{3}$ & Attractor & Not Physically Accepted\\
$Q_{B}^{1-}$ & Saddle & $B_{0}$\\
$Q_{B}^{1-}$ & Saddle & Big Rip\\
$Q_{B}^{2+}$ & Attractor & $B_{0}$\\
$Q_{B}^{2-}$ & Source & Big Rip\\
$Q_{B}^{3\pm}$ & Saddle & $B_{0}\left(  t-t_{0}\right)  ^{\frac{2}{3}}$\\
$Q_{B}^{4}$ & Saddle & $B_{0}\left(  t-t_{0}\right)  ^{\frac{3}{2}}$\\
$Q_{B}^{5}$ & Saddle & $%
\begin{array}
[c]{c}%
B_{0}~,~\frac{\sqrt{2}}{2}<Y<1\\
\text{Big Rip~,~}-1<Y\leq~\frac{\sqrt{2}}{2}%
\end{array}
$\\\hline\hline
\end{tabular}
\end{table}%

\subsection{Case C: $~A^{\prime}\neq0,~\gamma\neq0,~\delta\neq0$}

In the cosmological scenario, the associated cosmological constant plays an important role and charge is not considered. The stationary points $P_{C}=\left(  \Omega_{\alpha}\left(  P\right)  ,\Omega_{\beta}\left(P\right)  ,\Omega_{\gamma}\left(P\right)  ,\Omega_{\delta}\left(  P\right) \right)  ~$of the dynamical system (\ref{c.01}), (\ref{c.02}), (\ref{c.03})
and (\ref{c.04}) are%

\begin{align*}
P_{C}^{1} &  =\left(  0,0,0,0\right)  ~,\\
P_{C}^{2} &  =\left(  \Omega_{\alpha}^{0},-1-\Omega_{\alpha}^{0},0,0\right) ~,\\
P_{C}^{3} &  =\left(  0,0,1,0\right)  ~,\\
P_{C}^{4} &  =\left(  0,0,0,-3\right)  ~.
\end{align*}

$P_{C}^{4}$ is the new stationary point which describes a spacetime dominated by a negative value of the cosmological constant. From the master equation we derive $H\left(  P_{C}^{4}\right)  =H_{0}$, that is, $a\left(  t\right) = a_{0}e^{H_{0}t}$. We observe that the solution decays when $H_{0}<0$.
\ Recall that $P_{C}^{3}$ is not physically accepted because $\Omega_{\gamma}\leq0$.

Regarding the eigenvalues of the stationary points, they are $P_{C}^{1}:\left\{  3,-1,1,\frac{1}{2}\right\}  ,~P_{C}^{2}:\left\{  -2,-1-\frac{\Omega_{\alpha}^{0}}{2},0,2\right\}  $ and $P_{C}^{4}=\left(-1,-2,-3,4\right)  $. We conclude that $P_{C}^{4}$ is the unique attractor in the finite regime, while $P_{C}^{1}$ and $P_{C}^{2}$ are saddle points and hence are unstable solutions. 

The existence and results of the stability analysis for the stationary point $P_{C}^{4}$ are not unexpected. The point describes a rapid exponential expansion (or decay for $H_{0}<0$) for the three-dimensional hypersurface, in agreement with the well-known result for the anisotropic spacetimes presented in \cite{wald}.

\subsubsection{Poincar\'{e} Variables}

We introduce the Poincar\'{e} map

\begin{align}
\Omega_{a} &  =\frac{X}{\sqrt{1-X^{2}-Y^{2}-Z^{2}-W^{2}}},~\Omega_{\beta}=\frac{Y}{\sqrt{1-X^{2}-Y^{2}-Z^{2}-W^{2}}}\text{,~}\\
\Omega_{\gamma} &  =\frac{Z}{\sqrt{1-X^{2}-Y^{2}-Z^{2}-W^{2}}},~~\Omega_{\delta}=\frac{W}{\sqrt{1-X^{2}-Y^{2}-Z^{2}-W^{2}}}\text{,~}\\
d\tau &  =\sqrt{1-X^{2}-Y^{2}-Z^{2}-W^{2}}dT,
\end{align}

and we investigate the stationary points at infinity, that is, on the surface $1-X^{2}-Y^{2}-Z^{2}-W^{2}=0$. 

We obtain the stationary points%

\begin{align*}
Q_{C}^{1\pm} &  =\left(  \pm1,0,0\right)  ~,\\
Q_{C}^{2\pm} &  =\left(  0,\pm1,0\right)  ~,\\
Q_{C}^{3\pm} &  =\pm\frac{\sqrt{2}}{2}\left(  1,-1,0\right)  ~,\\
Q_{C}^{4} &  =\frac{1}{2}\left(  2X_{0},-X_{0}+\sqrt{2-3X_{0}^{2}}%
,-X_{0}-\sqrt{2-3X_{0}^{2}}\right)  ~,\\
Q_{C}^{5} &  =\left(  0,Y_{0},-\sqrt{1-Y_{0}^{2}},0\right)  ~,\\
Q_{C}^{6\pm} &  =\frac{1}{2}\left(  2X_{0},,-X_{0}\pm\sqrt{2-3X_{0}^{2}%
},0,-\left(  X_{0}\pm\sqrt{2-3X_{0}^{2}}\right)  \right) ~, \\
Q_{C}^{7\pm} &  =\left(  0,0,0,\pm1\right)~.
\end{align*}

The new stationary points are the $Q_{C}^{6\pm}$ and $Q_{C}^{7\pm}$.
$Q_{C}^{7\pm}$ describe solutions dominated by the cosmological constant.
Indeed, point $Q_{C}^{7+}$ provides $H\left(  Q_{C}^{7+}\right)  \rightarrow
0$, that is, $B\left(  Q_{C}^{7+}\right)  \rightarrow B_{0}$, while for point
$Q_{C}^{7-}$ we calculate, $B\left(  Q_{C}^{7-}\right)  \rightarrow+\infty$.
On the other hand, the stationary points provide $H\left(  Q_{C}^{6\pm
}\right)  =\frac{2}{3\left(  t-t_{0}\right)  }$, that is, $B\left(
Q_{C}^{6\pm}\right)  =B_{0}\left(  t-t_{0}\right)  ^{\frac{2}{3}}$. As far as
the stability is concerned, $Q_{C}^{6\pm}$ and $Q_{C}^{7\pm}$ are saddle
points while the rest of the points keep the same stability properties as
before. 

\begin{figure}[ptbh]
\centering\includegraphics[width=0.5\textwidth]{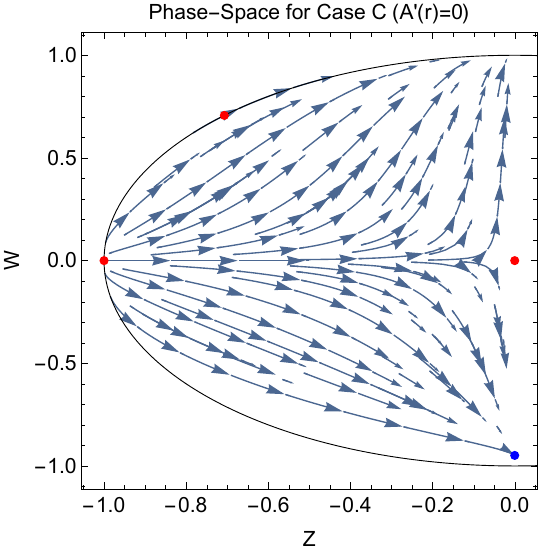}\caption{Phase-Space
for the dynamical system  (\ref{c.01}), (\ref{c.02}), (\ref{c.03}) and
(\ref{c.04}) in the compactified variables, on the surface where $A^{\prime
}\left(  r\right)  =0$, that is, $\Omega_{\alpha}=0$ and $\Omega_{\beta}=0$.
The unstable asymptotic solutions are marked with red points, the attractor
$P_{C}^{4}$ is marked with blue. }%
\label{fig2}%
\end{figure}

The results are summarized in Table \ref{table3}. In Fig. \ref{fig2} we
present the phase-space of the (\ref{c.01}), (\ref{c.02}), (\ref{c.03}) and
(\ref{c.04}) in the compactified variables, on the surface where $A^{\prime
}\left(  r\right)  =0$, that is, $\Omega_{\alpha}=0$ and $\Omega_{\beta}=0$. 

\begin{table}[tbp] \centering
\caption{Case C: Stationary points and stability properties}%
\begin{tabular}
[c]{ccc}\hline\hline
\textbf{Point} & \textbf{Stability} & $\mathbf{B}\left(  t\right)  $\\\hline
$P_{D}^{1}$ & Source & $B_{0}\left(  t-t_{0}\right)  ^{\frac{2}{3}}$\\
$P_{D}^{2}$ & Saddle & $B_{0}\left(  t-t_{0}\right)  $\\
$P_{D}^{3}$ & Saddle & Not Physically Accepted\\
$P_{D}^{4}$ & Attractor & $B_{0}e^{H_{0}t}$\\
$Q_{D}^{1-}$ & Saddle & $B_{0}$\\
$Q_{D}^{1-}$ & Saddle & Big Rip\\
$Q_{D}^{2+}$ & Attractor & $B_{0}$\\
$Q_{D}^{2-}$ & Source & Big Rip\\
$Q_{D}^{3\pm}$ & Saddle & $B_{0}\left(  t-t_{0}\right)  ^{\frac{2}{3}}$\\
$Q_{D}^{4}$ & Saddle & $B_{0}\left(  t-t_{0}\right)  ^{\frac{3}{2}}$\\
$Q_{D}^{5}$ & Saddle & $%
\begin{array}
[c]{c}%
B_{0}~,~\frac{\sqrt{2}}{2}<Y<1\\
\text{Big Rip~,~}-1<Y\leq~\frac{\sqrt{2}}{2}%
\end{array}
$\\
$Q_{C}^{6\pm}$ & Saddle & $B_{0}\left(  t-t_{0}\right)  ^{\frac{3}{2}}$\\
$Q_{C}^{7\pm}$ & Saddle & $%
\begin{array}
[c]{c}%
Q_{C}^{7+}:B_{0}\\
Q_{C}^{7-}:\text{Big Rip}%
\end{array}
$\\\hline\hline
\end{tabular}
\label{table3}%
\end{table}%

\section{Conclusion}

A phase plane analysis of the boundary condition for self-gravitating, radiating collapse has been undertaken. The associated nonlinear, second-order differential equation has been a topic of continued research since its inception by Santos \cite{san} and application by de Oliviera et al \cite{oli}. Initially, a solution was obtainable using a shear-free metric with the assumption of variable separable metric functions. 

In our study, we have a more general situation which includes the Einstein-Maxwell system for charge inclusion and also a cosmological constant. This complicates the boundary condition considerably, making it more difficult to obtain solutions. Reduction to a first-order system is possible by applying a suitable transformation which has been done in our study. This then leads to the phase-plane analysis in search of viable solutions.

Three cases have been considered, namely, a neutral fluid, a charged fluid and a charged fluid within a non-zero cosmological setting. For a neutral fluid, we see that there is an attractor at infinity and we also conclude that a radiating system would continue to collapse indefinitely without horizon formation. In the case of a charged fluid, similar results were obtained as in the neutral fluid case. Additional, asymptotic solutions were obtained due to a non-zero charge, however, these corresponded to saddle points and are hence unstable. In the most general case that we studied involving a non-zero cosmological constant, further solutions were obtained but again these were saddle points suggesting instability. In particular, an attractor was obtained which was associated with exponential time dependence for the metric function $B(t)$. {The latter correspond to the cosmological background where the object lies.}

We believe that our results support the recent study of Maharaj and Govinder \cite{mah}, and are an extension of their analysis. In \cite{mah} they gave well-tabulated physical outcomes of the various situations according to the parameter restrictions. Moreover, in our study we determine the asymptotic behaviour for the scale factor in the stationary limit. This provides us with important information about the time-evolution of the non-adiabatic collapse of the fluid. 

In future work, we plan to employ this approach in a more general geometric scenario with non-zero vorticity. 
\\

\label{sec4}

\textbf{Data Availability Statements:} Data sharing is not applicable to this
article as no datasets were generated or analyzed during the current study.
\newline\newline\textbf{Code Availability Statements:} Code sharing is
available after request. \newline

\textbf{Author Contribution Declaration: }AP initiated the project, RB abd AP performed the main analysis
and wrote the first draft. GL verified the analysis, produced the plots and the tables. MG, KS and SM wrote the 
discussion in the study. All authors reviewed the manuscript.

\begin{acknowledgments}
GL \& AP were supported by Proyecto Fondecyt Regular 2024, Folio 1240514,
Etapa 2025. The authors thank the organizers of the  Workshop "Nonlinear Systems" , which took place on November 12th at Durban, where this collaboration was initiated. 
\end{acknowledgments}

\end{document}